\documentclass[
    ,final            
  ]
  {aipproc}
  \pdfoutput=1

\layoutstyle{6x9}
\usepackage{graphicx}

\newcommand{\wt}[1]{\widetilde{#1}}
\newcommand{\bp}[1]{\left(#1\right)}
\newcommand{\bc}[1]{\left\{#1\right\}}
\newcommand{\bb}[1]{\left[#1\right]}
\newcommand{\Ochi}{\Omega_\chi {\rm h}^2}

\begin{document}

\title{Dark matter signals in deflected mirage mediation}
 
\classification{12.60.Jv,14.80.Ly,95.35.+d}

\keywords{Dark matter theory, dark matter experiments, supersymmetry breaking}
 
\author{Michael Holmes}{
  address={Department of Physics, Northeastern University, Boston, MA
02115, USA}
}

\begin{abstract}
We investigate the parameter space of a specific class of model within the deflected mirage mediation (DMM) scenario.  We look at neutralino properties and compute the thermal relic density as well as interaction rates with xenon direct detection experiments.  We find that there are portions of the parameter space which are in line with the current WMAP constraints.  Further we find that none of the investigated parameter space is in conflict with current bounds from the Xenon10 experiment and that future large-scale liquid xenon experiments will probe a large portion of the model space. 

\end{abstract}

\maketitle


Typically one imagines a world in which supersymmetry breaking is mediated via one of the three typcial scenarios, i.e. supergravity effects, a super-conformal anomaly or through gauge mediation.  Recently in \cite{Holmes:2009mx} we investigated the neutralino properties and dark matter signals within the framework of the deflected mirage mediation (DMM) scenario \cite{Everett:2008qy,Everett:2008ey,Choi:2009jn}.  In the DMM scenario all three mediation mechanisms are present and may be of similar size.  These models are generalizations of the mirage mediation scenario (see \cite{Choi:2007ka} and references therein) of which KKLT type models are an example.  
\par To outline the DMM model setup we briefly discuss the parameters and sketch the resulting low scale gaugino masses at one-loop.  First at some high scale which we take to be the GUT scale, $\mu_{\rm GUT}\simeq2\times 10^{16}$ GeV, the soft terms receive contributions from both the Planck-suppressed operators ($M_0$) as well as from the superconformal anomaly ($m_{3/2}$).  The presence of gauge mediation introduces a contribution to the soft terms ($\Lambda_{\rm mess}$) at some intermediate scale $\mu_{\rm mess}<\mu_{\rm GUT}$ due to $N_m$ guage charged messengers running in the loops. The messengers are taken in complete GUT representations to preserve gauge coupling unification i.e. $1/g_a^2\bp{\mu_{\rm GUT}}=1/g^2_{\rm GUT}-N_m\ln\bp{\mu_{\rm GUT}/\mu_{\rm mess}}/8\pi^2$ with $g^2_{\rm GUT}\simeq0.5$.  The presence of the guage charged messgengers also alters the beta function coefficients above $\mu_{\rm mess}$ where $b_a'=b_a+N_m$ with $b_a=\{33/5,1,-3\}$ for the MSSM.  
\par In the DMM scenario the three mass contributions to the soft terms are collected as one overall mass scale $M_0$ and two dimensionless ratios $\alpha_m=m_{3/2}/\left( M_0\ln(M_{\rm PL}/m_{3/2})\right)$ and $\alpha_g=\Lambda_{\rm mess}/m_{3/2}$ with the reduced Planck mass $M_{\rm PL}=2.4\times10^{18}$ GeV included as in the mirage literature.  Following Choi \cite{Choi:2009jn} it is convenient to introduce dimensionless variables $x$ and $y$  
\begin{equation} 
x=1/\bp{R+\alpha_m}\, , \quad\quad\quad
y=\alpha_m/\bp{R+\alpha_m}\, , \label{xydef} 
\end{equation}
where the dimensionless quantity $R$ is 
\begin{equation} 
R = 1 - \frac{N_m g^2_{\rm GUT}}{8\pi^2}\bc{\frac{\alpha_m\alpha_g}{2}\ln\bp{\frac{M_{\rm PL}}{m_{3/2}}}+\ln\bp{\frac{\mu_{\rm GUT}}{\mu_{\rm mess}}}} \,.
\end{equation} 
It is usefull to display results in the $xy$ plane for a fixed $M_0$, as the various limits of the theory may be convienently reached (see \cite{Choi:2009jn}); for example the line $x+y=1$ represents a mirage model. 
The net effect on the gaugino masses of running to some low scale $\mu<\mu_{\rm mess}$ may be written in terms of a set $\{M_0,x,y\}$ along with fixed values of $N_m$ and $\mu_{\rm mess}$ as
\begin{equation}
M_a(\mu) =M_0\bp{1+\beta_a(\mu)t} x^{-1} \bc{1+y\bb{\frac{\beta_a(\mu)t'}{1+\beta_a(\mu)t} -1}} \, , \label{Ma low}
\end{equation}
where $t=\ln\bp{\mu/\mu_{\rm GUT}}$, $t'=0.5\ln\bp{M_{\rm PL}/m_{3/2}}$ and $\beta_a(\mu)=b_ag_a^2(\mu)/8\pi^2$.
\par The soft gaugino masses at the low scale are not enough to fully specify the gaugino sector of the theory, one also needs the supersymmetric Higgs mass $\mu$.  One computes the low scale soft scalar masses $m_i^2$ and the trilinear couplings $A_{ijk}$ (full expressions may be found in the DMM literature) and determines $\mu$ via the EWSB conditions.  Both the soft scalar masses and the trilinear couplings depend on the modular weights $n_i$ of the matter superfields in the theory. 
\par
\begin{figure}[t]
    \includegraphics[scale=0.5]{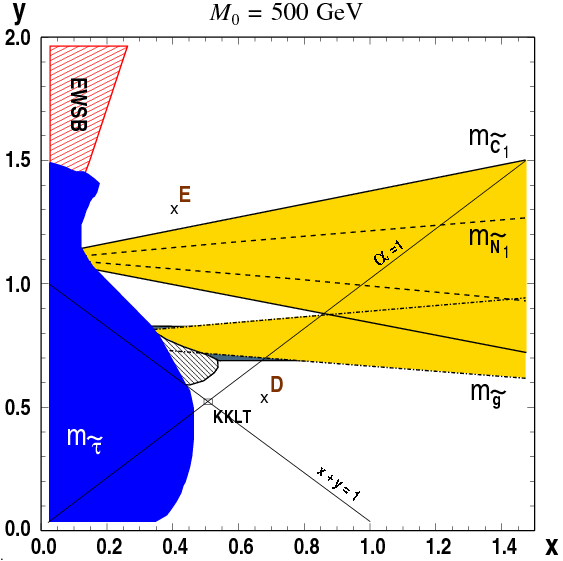}
    \includegraphics[scale=0.5]{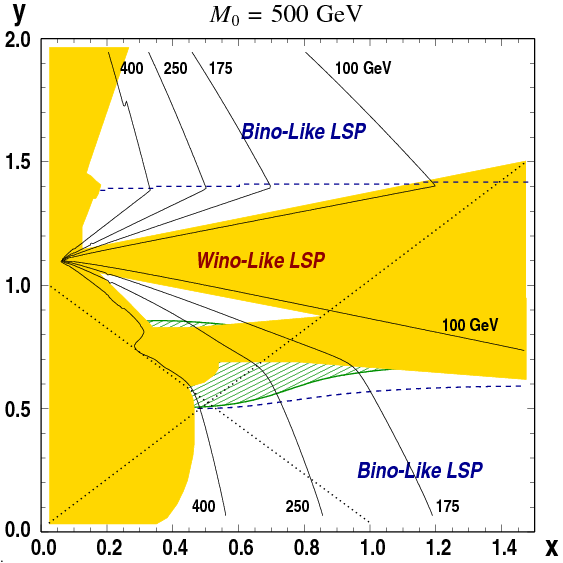}
  \caption{\label{fig:explicit model LSP's and wvfnc}{\footnotesize Parameter space properties. See text for details.}}
\end{figure}

 Having outlined the framework and necessary ingredients of a DMM model we now choose an explicit model class and investigate the model properties relevant to the thermal relic density and direct detection.  First we choose the modular weights for the matter and Higgs fields to be $n_m=1/2$ for $m=Q,u,d,L,e$ and $n_H=1$ for $H=H_u,H_d$.  To finish specifying the model we take $\{M_0,\mu_{\rm mess}\}=\{500,10^{10}\}$ GeV and $\{N_m,\tan\beta\}=\{3,10\}$ and show some resulting features in the $xy$ plane in Figure \ref{fig:explicit model LSP's and wvfnc}.  
In the left panel we show the regions of non-$\wt{N}_1$ LSP's, areas where EWSB conditions are not satisfied and bounds for the $\wt{N}_1$, $\wt{C}_1$ and $\wt{g}$ masses.  The blue region has a $\wt{\tau}$ LSP and the small hatched region has a $\wt{t}$ LSP.  The upper red region does not break EWSB, i.e. $m_A^2<0$.  In the yellow region we outline regions where masses are below current accelerator bounds of $m_{\wt{N}_1}<46$ GeV, $m_{\wt{C}_1}<103$  GeV and a conservative bound of $m_{\wt{g}}<200$ GeV.  The right panel gives the contours of constant $\wt{N}_1$ mass as well as the wavefunction composition, labelled as bino-like where the $\wt{N}_1$ is over 95\% bino, etc. and  the hatched region has a sizable Higgsino content. 
\par
\begin{figure}[b]
    \includegraphics[scale=0.535]{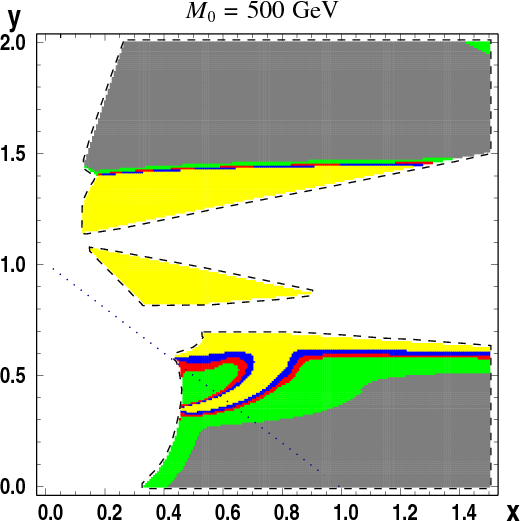}
    \includegraphics[scale=0.5]{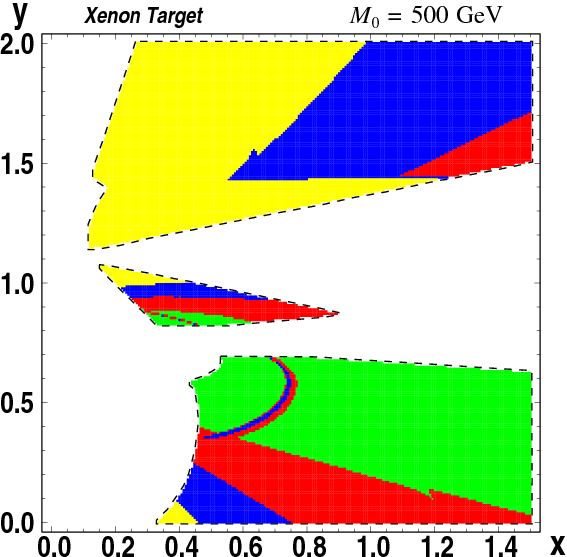}
  \caption{\label{fig:explicit model relic and dd rates}{\footnotesize $\Ochi$ and direct detection rates on xenon. See text for details.}}
\end{figure}

  We now examine the thermal relic density and some direct detection properties which are summarized in Figure \ref{fig:explicit model relic and dd rates}.  All calculations are done using the numerical code \texttt{DarkSUSY} \cite{Gondolo:2004sc}.  The three-year WMAP data~\cite{Spergel:2006hy} gives
\begin{equation}
0.0855 \leq \Omega_\chi {\rm h}^2 \leq 0.1189 \label{omegah2}
\end{equation}
at the $2\sigma$ level. The left panel of Figure \ref{fig:explicit model relic and dd rates} shows the thermal relic density from a scan of the $xy$ plane.  We show regions in red which are favorable to the WMAP constraints with $0.07<\Ochi<0.14$.  The blue regions have $0.025<\Ochi<0.07$ and the yellow regions are below the critical density with $\Ochi<0.025$ where we have wino LSP's or an $A$-funnel region where $m_A/2\simeq m_{\wt{N}_1}$.  In the green and gray regions we have $0.14<\Ochi<1$ and $1<\Ochi$, repectively.  We note that the standard relic density computation does not account for non-thermal production which may increase the abundance for wino models, and there are also mechanisms which exist that may bring over abundant regions in line with WMAP. 
\par In the right panel of Figure \ref{fig:explicit model relic and dd rates} we show estimates of direct detection rates on a liquid xenon target which is typical for both current and planned experiments \cite{Angle:2007uj,Aprile:2004ey,LUX}.  As in \cite{Altunkaynak:2008ry} we compute the rate of interactions of neutralinos with nuclei over a recoil energy range relevant for xenon experiments of 5 to 25 keV.  Note that in regions where $\Ochi<0.025$ we rescale the local halo density in the rate calculations.  Expected backgrounds for this type of experiment are on the order of 10 (or less) events per year of exposure.  To estimate interaction with experiments we compute the rate $R_{10}$ needed for relevant xenon experiments to have 10 events in a given exposure as summarized in Table \ref{table:ddrates}.  
\begin{table}
\begin{tabular}{l||c|c||c} 
 Experiment Name & Fiducial Mass [kg] & Exposure Time [yr]
& $R_{10}$ [counts/(kg yr)] \\
\hline
 XENON10 & 5.4 & 0.16 & 11.54 \\
 XENON100 & 170 $\times$ 0.8 & 1 & $7.35\times 10^{-2}$ \\
 LUX & 350 $\times$ 0.8 & 3 & $1.19\times 10^{-2}$ \\
 XENON1T & 1000 $\times $0.8 & 5 & $2.50\times 10^{-3}$ \\
\end{tabular}
{\caption{\label{table:ddrates}\footnotesize {\bf Rate Estimates for
Various Experiments}. See text for details.}}
\end{table} 
Note that for Xenon10 we use the experimentally quoted fiducial mass, while for future experiments we assume a fiducial mass equivalent to 80\% of the nominal quoted target mass.  We divide the $xy$ plane into regions based on $R_{10}$ for the xenon experiments listed in Table \ref{table:ddrates}.  For the models considered here we do not find that any would have given a signal over background at the Xenon10 experiment.  The green shading indicates models which will be probed at roughly the 100 kg-year level with $9.18>R_{10}\geq 0.0735$ counts/kg-yr.  At about the one ton-year level the red regions with $0.0735>R_{10}\geq 0.0119$ counts/kg-yr will be probed.  The blue region will be probed after about 5 years of Xenon1T and has $0.0119>R_{10}\geq 0.0025$ counts/kg-yr.  Finally regions in yellow are inaccessible to experiments considered in Table \ref{table:ddrates}.

\subsubsection{Acknowledgements}
This work is supported by the National Science
Foundation under the grant PHY-0653587.

\end{document}